\documentclass[11pt]{article}
\usepackage{amsmath}
\usepackage{amssymb}
\usepackage{latexsym}
\usepackage{epsfig}

\newcommand{\dt}{t}
\newcommand{\csf}{\Psi}
\newcommand{\G}{G}
\newcommand{\HH}{H}
\newcommand{\be}{\begin{equation}}
\newcommand{\ee}{\end{equation}}

\newcommand{\sectiono}[1]{\section{#1}\setcounter{equation}{0}}

\newcommand{\p}{\partial}

\begin{document}

{}~
\hfill\vbox{\hbox{hep-th/0409018}
\hbox{IFT-P.041/2004}\hbox{MIT-CTP-3537}
}\break

\vskip 3.0cm

\centerline{\Large \bf WZW-like Action for
Heterotic String Field Theory}

\vspace*{10.0ex}

\centerline{\large Nathan Berkovits${}^1$, Yuji Okawa${}^2$,
and Barton Zwiebach${}^2$}

\vspace*{7.0ex}

\centerline{\large \it ${}^1$ Instituto de F\'{\i}sica Te\'orica}

\centerline{\large \it
Universidade Estadual Paulista}

\centerline{\large \it R. Pamplona 145, S\~ao Paulo, SP 01405-900, BRASIL}
\vspace*{1.0ex}

\centerline{nberkovi@ift.unesp.br}

\vspace*{4.0ex}

\centerline{\large \it ${}^2$ Center for Theoretical Physics}

\centerline{\large \it
Massachusetts Institute of Technology}

\centerline{\large \it Cambridge,
MA 02139, USA}
\vspace*{1.0ex}

\centerline{okawa@lns.mit.edu, zwiebach@lns.mit.edu}

\vspace*{10.0ex}

\centerline{\bf Abstract}
\bigskip
\smallskip

We complete the construction of the Neveu-Schwarz sector
of heterotic string field
theory begun in hep-th/0406212 by giving a closed-form expression for the
action and gauge transformations. Just as the
Wess-Zumino-Witten (WZW) action for open superstring field theory can
be constructed from pure-gauge fields in bosonic open string field theory,
our heterotic string field theory action is constructed from pure-gauge
fields in bosonic closed string field theory. The construction involves
a simple alternative form of the WZW action which is consistent with the
algebraic structures of closed string field theory.

\vfill \eject

\baselineskip=16pt

\tableofcontents

\sectiono{Introduction}

In bosonic open string field theory, the first-quantized condition $Q A=0$
for physical vertex operators is generalized to the nonlinear equation of
motion
\begin{equation}
\label{bosoneom}
Q A + A A =0
\end{equation}
where string fields are multiplied
using the open string star product~\cite{Witten:1985cc}.
To extend
this result to the
Neveu-Schwarz (NS) sector of open superstring field theory,
an
equation of motion $Q A + Z( A A) =0$ was proposed where $Z$
is a picture-raising operator inserted at the string
midpoint~\cite{Witten:1986qs}.
Approaches involving midpoint picture-changing insertions, however,
have contact-term problems which lead to a breakdown of gauge
invariance~\cite{Wendt:1987zh}.

This difficulty is resolved in open superstring field theory by
working in the `large' Hilbert space of Friedan, Martinec, and
Shenker~\cite{Friedan:1985ge},
     which includes the $\xi$ zero mode coming from bosonization of the
$(\beta,\gamma)$ ghosts. Vertex operators $\Phi$
in the large Hilbert space are related to vertex operators $A$
in the small Hilbert space by $\eta \Phi = A$ where
$\eta$ denotes the zero mode of the field  conjugate to
$\xi$.  Therefore, the first-quantized condition $QA=0$ implies that
$\eta Q\Phi=0$ is the linearized equation of motion in the large
Hilbert space.  The operators $\eta$ and $Q$ anticommute and both
of them square to zero.

The linearized equation for $\Phi$ generalizes to the nonlinear
equation of motion
\begin{equation}
\label{osf}
\eta (e^{-\Phi} Q e^{\Phi})=0,
\end{equation}
which can be derived from a Wess-Zumino-Witten (WZW)
action for
open superstring field theory which does not involve explicit
picture-changing operator insertions~\cite{Berkovits:1995ab}.
Note that the open superstring equation of motion of
(\ref{osf}) can
be written as $\eta \overline A_Q=0$ where $\overline A_Q=e^{-\Phi} Q
e^\Phi$ is
in fact a pure-gauge solution of the {\em bosonic} open string field theory
equation of motion of (\ref{bosoneom}).

In bosonic closed string field theory,
the first-quantized condition for physical vertex
operators $\Psi$ also takes
the form $Q\Psi=0$, where
$Q = Q_L+Q_R$, is the BRST operator obtained by adding the
contributions from the left-moving $(L)$ and right-moving $(R)$ sectors.
This equation
is generalized to the following nonlinear equation of
motion
\begin{equation}
\label{bsf}
Q\Psi + \sum_{n=2}^\infty
{\kappa^{n-1}\over{n!}}[\Psi^n]=0\,,
\end{equation}
where string fields are multiplied using
the various closed string products~\cite{Zwiebach:1992ie,Saadi:tb}
and $\kappa$ is the gravitational constant.
Although one can extend this equation to heterotic
strings by non-canonical insertions of
picture-changing operators~\cite{Saroja:1992vw},
it was recently shown by two of the authors
that insertions of picture-changing operators can be
altogether avoided in heterotic
string field theory if one works in the large Hilbert
space~\cite{Okawa:2004ii}.
Denoting by $\eta = \eta_L$ the zero mode
of the left-moving superghost,  the first-quantized condition
for NS heterotic vertex operators $V$
in the large Hilbert space takes the form
$\eta Q V=0$.  The  authors constructed the first few terms
in the nonlinear generalization
of the heterotic equation of motion, action,
and gauge transformations.

In this paper, we complete this construction by giving closed-form
expressions for the complete equation of motion, action,
and gauge transformations in the NS
sector of heterotic string field theory. Just as $\eta \overline A_Q=0$
is the open superstring equation of motion where $\overline A_Q$ is a
pure-gauge solution to bosonic open string field theory, we show that
$\eta\,\overline \csf_Q=0$ is the heterotic equation of motion
where $\overline \csf_Q$ is a pure-gauge solution to the {\it bosonic}
closed string equation of motion of (\ref{bsf}).
The heterotic string field theory action takes
a rather simple but apparently unfamiliar form of the WZW action.

The familiar form of the WZW action includes two terms: one is written
in two-dimensional covariant language and the other,
the WZ term, is elegantly written as a  three-dimensional
covariant  expression~\cite{Witten:1983ar}.  In the form we have
found, the action is
given by just one term.  This term involves
integration over three dimensions, just like the WZ term but,
as written, is not manifestly two-dimensional nor three-dimensional
covariant.
Both for ordinary bosonic WZW models and for open superstring
theory,
this alternative form of the action is completely equivalent
to the more familiar one.
For the heterotic string, the non-standard form of the WZW action
appears to provide the fundamental definition of the theory.  The
transformation to the familiar WZW action is not generally valid,
although it is possible in the case where the string field
is chosen to have `linear homotopy' in the third dimension.

As we explain in detail, our construction uses the BRST operator
$Q$ and three more operators: the zero mode $\eta$ of the superghost,
the derivative $\partial_t$ along the third direction of WZW theory,
and the variation $\delta$.  In open superstring field theory
these
four operators are all derivations of
the star algebra of open strings. In particular, $Q$ and $\eta$
appear in the action symmetrically.  On the other hand, the four
operators are not
on the same footing in heterotic theory: $Q$ is not a derivation of
all closed string
products, while the other three are.  In particular, $Q$ and $\eta$
play different roles and do not appear symmetrically in the
action.

\medskip
In section 2 of this paper we rewrite the WZW action for
open superstring field theory in the
non-standard form which generalizes to heterotic string field
theory. In section 3
we use this non-standard form of the WZW action to
construct an explicit action for heterotic string field theory and show that
it implies the desired equation of motion and gauge invariances. We
also show that the first few terms in the expansion of the action coincides
with the terms constructed in \cite{Okawa:2004ii}. Some
of the structures in open string field theory are absent in closed string
field theory, and the non-standard form of the WZW action does not require
these additional structures. In section 4 we discuss the relationship
of our heterotic action with the more familiar form for the WZW action.
Finally, in section 5 we summarize our results and discuss some possible
applications.

\sectiono{Open superstring field theory}

In this section we show that the familiar action
for WZW theory
can be recast in a rather useful and simple way.  We then exhibit
a similar recasting for the WZW open superstring field theory action.
Our construction of the heterotic string field
theory action will be based on an analogous simple form.
The analysis of open superstrings in this section will use the tools
that generalize to heterotic strings.

\subsection{Rewriting the WZW action}

Consider bosonic WZW theory on a two-dimensional compact
space with coordinates $z$ and $\bar z$.  As usual, to write
a natural action we introduce a three-dimensional space.  The
extra dimension can be parametrized by a coordinate $t\in [0,1]$
and we introduce a Lie-algebra valued field $\Phi (t, z, \bar z)$.
The field $\Phi$ is required to vanish
at $t=0$ for all $z$ and $\bar z$.
The original two-dimensional
space is obtained for $t=1$.  The WZW action is written in terms
of a pure-gauge connection
\begin{equation}
A_i  = e^{-\Phi} \, (\partial_i e^\Phi )\,,   \qquad  i = t, z, \bar z\,.
\end{equation}
By construction, this connection is flat
\begin{equation}
F_{ij} = \partial_i A_j - \partial_j A_i
+ [\, A_i ,\, A_j \,] =0 \,,
\end{equation}
and both $A_z$ and $A_{\bar z}$ vanish for $t=0$.
The WZW action is now given as
\begin{equation}
\label{wzwconventional}
S ={1\over {2g^2}} \int d^2 z \int_0^1 dt\,\, \hbox{Tr} \Bigl(
\,\partial_t(A_z
\,  A_{\bar z})  +   A_t [ A_z, A_{\bar z}]~\Bigr)\,.
\end{equation}
The integral over $t$ can be explicitly done for the first term inside the
parentheses; it gives the familiar kinetic term defined on the original
two-dimensional space.
We now show that the above action can be rewritten as
\begin{equation}
\label{newf}
S = {1\over {g^2}}\int d^2 z \int_{0}^{1} dt \,\, \hbox{Tr}
\bigl( \, (\partial_{z} A_t)\, A_{\bar z} \bigr)\,.
\end{equation}
To prove this we begin with (\ref{wzwconventional}),
expand the $t$ derivatives and use the vanishing of $F_{tz}$ and
$F_{t\bar z}$ to trade the $t$ derivatives for $z$  and $\bar z$
derivatives:
\begin{equation}
S =
{1\over {2 g^2}} \int d^2 z \int_{0}^{1} dt \,\,
\hbox{Tr} \Bigl(\,(\p_z A_t)
A_{\bar z}  + A_z (\p_{\bar z} A_t) + A_t [ A_z, A_{\bar z}]\Bigr)\,.
\end{equation}
Integrating by parts the $\bar z$ derivative and using
$F_{z\bar z} =0$ we find
\begin{equation}
S =
{1\over {2g^2}} \int d^2 z \int_{0}^{1} dt \,\,
\hbox{Tr} \Bigl(\,(\p_z A_t)
A_{\bar z}  - (\p_z A_{\bar z}) A_t) \Bigr)
={1\over {g^2}} \int d^2 z \int_{0}^{1} dt
\,\, \hbox{Tr} \bigl(\,(\p_z A_t) A_{\bar z}
\bigr)\,,
\end{equation}
as we wanted to show.

\subsection{Open superstring field theory}

The WZW open superstring field theory action is given by
\begin{equation}
\label{wzwopenstring}
S=-\frac{1}{2g^2} \int_0^1 dt \,
\langle\langle \, \partial_t \, ( A_\eta \, A_Q )
+ A_\dt \,\{ A_Q \,,A_\eta\} \, \rangle\rangle \,,
\end{equation}
where $\{ A, B\} \equiv AB+ BA$. We also use  $[A, B ] \equiv  AB -
BA$, and more
generally,
\begin{equation}
\label{gcomm}
[\, A\,, B\,\} \equiv  AB - (-1)^{AB} BA \,,
\end{equation}
where string fields or operators in the exponent represent
their Grassmann property, $0$ (mod $2$) if Grassmann even
and $1$ (mod $2$) if Grassmann odd.
The double brackets in (\ref{wzwopenstring}) denote
correlators defined in the large Hilbert space
using the geometry of the star product and
its iterated versions, and
\begin{equation}
\label{nolabel}
A_Q = e^{-\Phi(t)} (Q e^{\Phi(t)}) \,, \quad
A_\eta = e^{-\Phi(t)} (\eta e^{\Phi(t)}) \,, \quad
A_\dt = e^{-\Phi(t)}  (\partial_t e^{\Phi(t)}) \,,
\end{equation}
where $\Phi (0) = 0$ and $\Phi (1) \equiv \Phi$.  Here
$\eta$ denotes the zero mode of the superghost field $\eta(z)$.
If we take $\Phi(t)=t \Phi$, these string fields become
\begin{equation}
\label{yuhiibgjfkjg}
A_Q = e^{-t\Phi} (Q e^{t\Phi}) \,, \quad
A_\eta = e^{-t\Phi} (\eta e^{t\Phi}) \,, \quad
A_\dt = e^{-t\Phi} (\partial_t e^{t\Phi}) = \Phi \,.
\end{equation}
The open superstring field theory action has been usually
written using this specific form of $\Phi(t)$.

The definitions in (\ref{nolabel}) are all of the form
\begin{equation}
A_X = e^{-\Phi(t)} ( X
e^{\Phi(t)})\,,
\end{equation}
where $X$ is a derivation of the star product of open string fields,
namely, $X$ satisfies
$X (AB) = (XA)B + (-1)^{XA} A(XB)$
for arbitrary string fields $A$ and $B$.\footnote{
We use $\dt$ for $\partial_t$ in indices.
For example, $A_\dt = A_{\partial_t}$.}
Indeed, $Q$, $\eta$, and $\partial_t$ are derivations.
They are also derivations of the graded commutator (\ref{gcomm}):
$X[A, B \} = [XA, B\}  + (-1)^{XA} [A, XB \}$.

The string field $A_Q$ has an
important interpretation.   It is a pure-gauge open string field
associated with
the `large' gauge parameter $\Phi(t)$ in bosonic open string field theory.
In this theory the equation of motion takes the form
$QA + AA =0$ and the infinitesimal gauge transformations take the form
$\delta_\epsilon A =  Q\epsilon + [\, A,\, \epsilon \,]$.
One can readily check that
$A_Q$ satisfies the string field equation of motion: $QA_Q + A_Q A_Q
=0$.

If we expand the bosonic open string field theory action about $A_Q$,
the new BRST operator $Q'$ in the action acts as follows:
\begin{equation}
\label{009ifgkjej}
Q' B = Q B + [\, A_Q ,\, B \,\} \,.
\end{equation}
The operator $Q'$ squares to zero,  is a derivation
of the star product, and is also a derivation of
the graded commutator. One can
view the $Q'$
action as that of a covariant derivative with gauge connection $A_Q$,
accordingly, we write
$\nabla_Q B \equiv Q' B$. The language of field strengths is quite natural
indeed. For any two derivations $X$ and $Y$ that commute or anticommute,
one finds that the field strength
$F_{XY}$ vanishes:
\begin{equation}
\label{algderos}
F_{XY} \equiv X A_Y -   (-1)^{XY} Y A_X + [A_X, A_Y\} =0 \,.
\end{equation}
We use this notation for field strengths even when $X$ or $Y$ is
the derivation $\delta$. For any
derivation
$X$ which commutes or anticommutes with
$Q$ the following useful identity holds:
\begin{equation}
X(Q' B) -(-1)^X Q'(XB) = [\, X A_Q ,\, B \,\} \,.
\label{Q'-X-commutator-open}
\end{equation}

As in the case of the ordinary WZW action,
the superstring field theory action (\ref{wzwopenstring}) can also be
written as
\begin{equation}
\label{ossfta}
S = -\frac{1}{g^2} \, \int_0^1 dt \,\langle\langle \,
( \eta \, A_\dt) \, A_Q \, \rangle\rangle \,.
\end{equation}
We will prove  the equivalence of the two actions later.
Note that $A_\eta$ no longer appears in (\ref{ossfta}).

The gauge invariance of (\ref{ossfta}) can be shown
by computing the general variation $\delta S$.
We can compute the variation of the integrand
using the explicit forms of $A_Q$ and $A_\dt$
to find
\begin{equation}
\label{vardensity}
\delta \, \langle\langle \,
(\eta A_\dt) \, A_Q \, \rangle\rangle
= \partial_t \, \langle\langle \, (\eta A_\delta ) \,
A_Q \, \rangle\rangle \,.
\end{equation}
Note that the variation of the string field is encoded in
$A_\delta$. This direct approach, however, is not available in
heterotic string field theory,
where the closed string fields
corresponding to $A_Q$ and $A_\dt$ do not have simple expressions.
Moreover, while all field strengths vanish for open superstrings,
the heterotic string analogs  do not all vanish.
It is therefore useful
to calculate $\delta S$
without relying on the explicit expressions of $A_Q$ and $A_\dt$
and to extract the minimum structures
which ensure gauge invariance of the action.
We find that gauge invariance of the action
follows from the following three equations:
\begin{eqnarray}
\partial_t A_Q &=& \phantom{-} \, Q' A_\dt \,,
\label{A_partial-equation}
\\
\delta A_Q &=& \phantom{-}\, Q' A_\delta \,,
\label{A_delta-equation}
\\
\eta A_Q &=& - \, Q' A_\eta \,.
\label{A_eta-equation}
\end{eqnarray}
These equations take the form $X A_Q = (-1)^X\nabla_Q A_X$
(which follows from $F_{XQ}=0$), for $X= \partial_t,
\delta$, and $\eta$.   In the computation of
$\delta S$ we also use the identity:
\begin{equation}
\label{theextraone}
Q' \, (\, \delta A_\dt - \partial_t A_\delta
+ [\, A_\delta ,\, A_\dt \,] \,) = 0 \,.
\end{equation}
This identity, which states that $Q'$ annihilates $F_{\delta \dt}$,
is not  independent.
It follows from
(\ref{A_partial-equation}), (\ref{A_delta-equation}), $[\delta\,,
\partial_t] =0$, and
(\ref{Q'-X-commutator-open}):
\begin{eqnarray}
0 &=& \delta ( \partial_\dt A_Q ) - \partial_t ( \delta  A_Q )
\nonumber \\
&=& \delta ( Q' A_\dt ) - \partial_t ( Q' A_\delta )
\nonumber \\
&=& Q'(\delta A_\dt) + [\, \delta A_Q ,\, A_\dt \,]
-Q'(\partial_t A_\delta) - [\, \partial_t A_Q ,\, A_\delta \,]
\nonumber \\
&=& Q'(\, \delta A_\dt - \partial_t A_\delta
+ [\, A_\delta ,\, A_\dt \,] \,) \,.
\end{eqnarray}

We are now ready to compute the variation of the action
(\ref{ossfta}).  The variation of the
integrand is
\begin{equation}
\delta \, \langle\langle \,
(\eta A_\dt) \, A_Q \, \rangle\rangle
= \langle\langle \,
(\eta \, \delta A_\dt) A_Q \, \rangle\rangle
+ \langle\langle \,
(\eta A_\dt) \, \delta A_Q \, \rangle\rangle \,.
\end{equation}
The first term on the right-hand side is evaluated as follows:
\begin{eqnarray}
\langle\langle \,
(\eta \, \delta A_\dt) \, A_Q \, \rangle\rangle
&=& - \langle\langle \,
\delta A_\dt \, (\eta A_Q) \, \rangle\rangle
= \langle\langle \,
\delta A_\dt \, (Q' A_\eta) \, \rangle\rangle
= - \langle\langle \,
(Q' \delta A_\dt) A_\eta \, \rangle\rangle
\nonumber \\
&=& - \langle\langle \, (\, Q' (\, \partial_t A_\delta
- [\, A_\delta ,\, A_\dt \,] \,) \,) \, A_\eta \,
\rangle\rangle
= \langle\langle \, (\, \partial_t A_\delta
- [\, A_\delta ,\, A_\dt \,] \,) \, (Q' A_\eta) \,
\rangle\rangle
\nonumber \\
&=& - \langle\langle \, (\, \partial_t A_\delta
- [\, A_\delta ,\, A_\dt \,] \,) \, (\eta A_Q) \,
\rangle\rangle
\nonumber \\
&=& \langle\langle \, (\, \eta \, \partial_t A_\delta \,) \,
A_Q \, \rangle\rangle
+ \langle\langle \, [\, A_\delta ,\, A_\dt \,] \,
(\eta A_Q) \, \rangle\rangle \,.
\end{eqnarray}
The second term on the right-hand side is evaluated as follows:
\begin{eqnarray}
\langle\langle \,
(\eta A_\dt) \, \delta A_Q \, \rangle\rangle
&=& \langle\langle \,
(\eta A_\dt) \, (Q' A_\delta) \, \rangle\rangle
= \langle\langle \,
( Q' (\eta A_\dt)) \, A_\delta \, \rangle\rangle
\nonumber \\
&=& - \langle\langle \,
( \eta \, (Q' A_\dt)) \, A_\delta \, \rangle\rangle
+ \langle\langle \,
[\, \eta A_Q ,\, A_\dt \,] \, A_\delta \, \rangle\rangle
\nonumber \\
&=& - \langle\langle \,
( \eta \, \partial_t A_Q ) \, A_\delta \, \rangle\rangle
+ \langle\langle \,
[\, \eta A_Q ,\, A_\dt \,] \, A_\delta \, \rangle\rangle
\nonumber \\
&=& - \langle\langle \,
\partial_t A_Q \, (\eta A_\delta) \, \rangle\rangle
+ \langle\langle \,
[\, \eta A_Q ,\, A_\dt \,] \, A_\delta \, \rangle\rangle
\nonumber \\
&=& \langle\langle \,
(\eta A_\delta) \, \partial_t A_Q \, \rangle\rangle
- \langle\langle \, [\, A_\delta ,\, A_\dt \,] \,
(\eta A_Q) \, \rangle\rangle \,.
\end{eqnarray}
Therefore, the variation of the integrand of the action
is given by (\ref{vardensity}).
Since the variation of the integrand of the action
is a total derivative in $t$, the variation of the action
is given by
\begin{equation}
\delta S=-{1\over g^2} \,  \,\langle \langle\,
(\eta\,\overline{A}_\delta)\,  \overline{A}_Q~ \rangle\rangle\,,
\end{equation}
where the bar denotes evaluation at $t=1$, namely,
$\overline{F} \equiv F(t=1)$.
Using (\ref{A_eta-equation}),
the variation of the action can also be written as
\begin{equation}
\delta S= {1\over g^2} \,  \,\langle \langle\,
(\overline{Q'} \,\overline{A}_\delta)\,
\overline{A}_\eta~ \rangle\rangle\,.
\end{equation}
Note that $A_\delta$ depends on $t$
only through $\Phi(t)$ and $\delta \Phi(t)$.
Therefore the relation between $A_\delta$ and $\delta \Phi(t)$ is invertible.
Since $\overline{Q'}\,{}^2$ and $\eta^2$ vanish, the action is invariant
under variations that satisfy
\begin{equation}
\overline{A}_\delta = \overline{Q'} \, \Lambda'
+ \eta \Omega \,.
\end{equation}
Therefore, we define gauge transformations
for $\Phi(t)$ by
\begin{equation}
\label{dflkgfdehr}
{A}_\delta = {Q'} \Lambda'(t) + \eta \,\Omega (t) \,,
\end{equation}
where $\Lambda' (0) = 0$, $\Lambda' (1) = \Lambda$,
$\Omega (0) = 0$, and $\Omega (1) = \Omega$.
Using the explicit forms of ${A}_\delta$
and ${A}_Q$, this equation can be written~as
\begin{eqnarray}
e^{-\Phi(t)} (\delta e^{\Phi(t)})
&=& Q \Lambda'(t) + e^{-\Phi(t)} (Q e^{\Phi(t)}) \, \Lambda'(t)
+ \Lambda'(t) \, e^{-\Phi(t)} (Q e^{\Phi(t)}) + \eta \Omega (t)
\nonumber \\
&=& e^{-\Phi(t)} \, (\, Q (e^{\Phi(t)} \Lambda'(t) e^{-\Phi(t)}) \,) \,
e^{\Phi(t)} + \eta \Omega (t) \,.
\end{eqnarray}
Therefore, we have
\begin{equation}
\delta e^{\Phi(t)} = (Q \Lambda(t)) \, e^{\Phi(t)}
+ e^{\Phi(t)} \, (\eta \,\Omega(t)) \,,
\end{equation}
where $\Lambda(t)= e^{\Phi(t)} \Lambda' (t)e^{-\Phi(t)}$ is a redefined
gauge parameter.
This is the usual form of the gauge transformations.

Let us next show the equivalence of this form of the action
and the WZW action (\ref{wzwopenstring}).
In the case of the ordinary WZW action,
we used three equations: $F_{tz}=0$, $F_{t \bar{z}}=0$,
and $F_{z \bar{z}}=0$.
Correspondingly, we also use three equations
in the case of open superstring field theory.
In addition to (\ref{A_partial-equation})
and (\ref{A_eta-equation}),
which correspond to $F_{\dt Q}=0$ and $F_{\eta Q}=0$,
respectively,
we need the vanishing of $F_{\dt \eta}$:
\begin{equation}
\label{dlkgoiiejnf}
\partial_t A_\eta - \eta A_\dt
+ [\, A_\dt ,\, A_\eta \,] = 0 \,.
\end{equation}
Using these three equations,
the integrand of the action (\ref{wzwopenstring})
can be rewritten in the following way:
\begin{eqnarray}
&& \frac{1}{2} \, \langle\langle \,
\partial_t \, ( A_\eta A_Q ) \, \rangle\rangle
+ \frac{1}{2} \, \langle\langle \,
A_\dt \, \{\, A_\eta ,\, A_Q \,\} \, \rangle\rangle
\nonumber \\
&=& \frac{1}{2} \, \langle\langle \,
(\, \partial_t A_\eta + [\, A_\dt ,\, A_\eta \,] \,) \,
A_Q \, \rangle\rangle
+ \frac{1}{2} \, \langle\langle \,
A_\eta \, ( \partial_t A_Q ) \, \rangle\rangle
\nonumber \\
&=& \frac{1}{2} \, \langle\langle \,
( \eta A_\dt ) \, A_Q \, \rangle\rangle
+ \frac{1}{2} \, \langle\langle \,
A_\eta \, ( Q' A_\dt ) \, \rangle\rangle
\nonumber \\
&=& \frac{1}{2} \, \langle\langle \,
( \eta A_\dt ) \, A_Q \, \rangle\rangle
- \frac{1}{2} \, \langle\langle \,
( \eta A_Q ) \, A_\dt \, \rangle\rangle
= \langle\langle \, ( \eta A_\dt ) \,
A_Q \, \rangle\rangle \,.
\end{eqnarray}
This completes the proof of equivalence.

We can view the construction of the action (\ref{ossfta})
in the following way.
The action  can be constructed from
$A_Q$, which is a pure-gauge solution
of bosonic open string field theory,
and $A_\dt$, which satisfies the relation (\ref{A_partial-equation}).
The action is gauge invariant because
$\delta A_Q$ and $\eta A_Q$ are $Q'$ exact, and
the gauge transformations (\ref{dflkgfdehr}) are consistent
because $A_\delta$  depends on $t$
only through $\Phi(t)$ and $\delta \Phi(t)$.
Such string fields $A_Q$, $A_\dt$, and $A_\delta$ can be constructed
by the following procedure.
First, take the pure-gauge string field
$e^{-\Phi} (Q \, e^{\Phi})$
and define $A_Q\equiv e^{-\Phi(t)} (Q \, e^{\Phi(t)})$,
where we have replaced $\Phi$ by $\Phi(t)$.
Since $A_Q$ is pure-gauge for arbitrary $\Phi (t)$,
its variation can be written
as an infinitesimal gauge transformation:
$\delta A_Q = Q' A_\delta$,
where $A_\delta$ is the gauge parameter.
Furthermore, $A_\delta$ is guaranteed to depend on $t$
only through $\Phi (t)$ and $\delta \Phi (t)$
because $A_Q$ and $Q'$ depend on $t$ only through $\Phi (t)$.
In  open superstring field theory,
such an $A_\delta$ can be found explicitly
and is given by $e^{-\Phi (t)} (\delta \, e^{\Phi (t)})$.
The verification that this expression for $A_\delta$
satisfies $\delta A_Q = Q' A_\delta$ relies
only on the derivation property of $\delta$
and the commutativity of $\delta$ and $Q$.
Since $\partial_t$ and $\eta$ are also derivations
that commute or anticommute with $Q$,
one can construct $A_\dt$ and $A_\eta$
satisfying (\ref{A_partial-equation}) and (\ref{A_eta-equation})
by simply replacing $\delta$ in $A_\delta$
with $\partial_t$ and $\eta$, respectively.
We will apply this strategy to heterotic string field theory
in the next section.

\sectiono{Heterotic string field theory}

In this section we first develop the closed string field theory
structures that are needed to construct the field theory
of heterotic strings~\cite{Gross:1985fr}.
As opposed to the open string theory
case, we have  no  concise
expression for a pure-gauge
closed string field. We  build the pure-gauge
closed string field with a finite gauge parameter $V$
by integrating infinitesimal gauge transformations.
The procedure we follow gives a pure-gauge closed string field
$\G (V)$
associated with $V$ that is
analogous to the open string
pure-gauge field $e^{-\Phi} (Q e^{\Phi})$ associated with $\Phi$.
We then define the closed string field
$\csf_Q \equiv \G (V(t))$,  just as
we defined $A_Q\equiv e^{-\Phi(t)} (Q e^{\Phi(t)})$.

Our next step is the  construction  of closed string fields
$\csf_\dt$, $\csf_\delta$, and
$\csf_\eta$ which satisfy
\begin{eqnarray}
\partial_t \csf_Q &=& \phantom{-} \, Q' \csf_\dt \,,
\label{E_partial-equation}
\\
\delta \csf_Q &=& \phantom{-} \, Q' \csf_\delta \,,
\label{E_delta-equation}
\\
\eta \csf_Q &=& - \, Q' \csf_\eta \,,
\label{E_eta-equation}
\end{eqnarray}
where $Q'$ is the BRST operator
of the bosonic closed string field theory action expanded
about $\csf_Q$.
The closed string fields
$\csf_\dt$, $\csf_\delta$, and $\csf_\eta$
are analogous to the open string fields
$A_\dt$, $A_\delta$, and $A_\eta$, respectively.
Finally, we propose the following string field theory action
\begin{equation}
S = {2\over {\alpha'}} \int_0^1 dt \,
\langle \, \eta \csf_\dt ,\, \csf_Q \, \rangle\,,
\label{hsfta}
\end{equation}
and we show that it has two gauge invariances, one generated by $Q$ and
one generated by $\eta$.
Moreover, when expanded in powers of the gravitational constant
$\kappa$, the action  (\ref{hsfta})  reproduces the kinetic term and
all the interaction terms
determined in \cite{Okawa:2004ii}.  The action (\ref{hsfta}) is a central
result in this paper.  It can be said to be a WZW-like action since it
takes a form analogous to the WZW action~(\ref{newf}).

\subsection{Shifted structures in closed string field theory}

\noindent
In closed string field theory the  algebraic structure includes
a bilinear form $\langle ~\cdot\,, \cdot ~\rangle$
which satisfies $\langle \, A,\, B \, \rangle
= (-1)^{(A+1) (B+1)} \, \langle \, B,\, A \, \rangle$,
a BRST operator $Q$, and a series of graded-commutative
string products $[\, B_1, \ldots , B_m \,]$
with $m \geq 2$
that satisfy a set of identities discussed in great detail in
\cite{Zwiebach:1992ie} and reviewed in \cite{Okawa:2004ii}.
The closed string field $\Psi$ is a ghost number two state of the
matter plus ghosts conformal field theory, and the (infinitesimal)
gauge parameter $\Xi$ is a state of ghost number one.
The action in bosonic closed string field theory is
\begin{equation}
S= -\frac{2}{\alpha'} \Bigl(
\frac{1}{2} \langle \, \Psi ,\, Q \Psi \, \rangle
+ \sum_{n=3}^\infty \frac{\kappa^{n-2}}{n!} \,
\langle \, \Psi ,\, [\, \Psi^{n-1} \,] \, \rangle \Bigr) \,,
\end{equation}
where $\kappa$ is the gravitational constant and we have
rescaled $\Psi$ so that the kinetic term has no
factor of  $1/\kappa^2$.
The gauge
transformation of bosonic closed string field theory is given by
\begin{equation}
\delta_{\,\Xi} \Psi = Q \, \Xi
+ \sum_{n=1}^\infty \frac{\kappa^n}{n!} \,
[\, \Psi^n ,\, \Xi \,] \,,
\end{equation}
and the equation of motion of the string field is
\begin{equation}
\label{stringfieldequation}
\mathcal{F} (\Psi) \equiv  Q\Psi + \sum_{n=2}^\infty
{\kappa^{n-1}\over n!} \, [\, \Psi^n\,] = 0 \,.
\end{equation}

It is useful to examine the closed string field theory
that arises when we expand the string field around a
nonvanishing string field $\csf_Q$.  To do this, we let
$\Psi \to \csf_Q + \Psi$ in the string field theory action
and reorganize the result in powers of $\Psi$.  This was done in
\cite{Zwiebach:1992ie} where a new
BRST-like operator $Q'$ and new products $[ \ldots\,]'$
were introduced:
\begin{equation}
\label{defshitedq}
Q' A  \equiv  QA + \sum_{n=1}^\infty {\kappa^n\over n!} \,
[\, \csf_Q^n \,, A \, ] \,,
\end{equation}
\begin{equation}
[\, B_1\, ,\, B_2 \,, \ldots, B_m \,]' \equiv \sum_{n=0}^\infty
\frac{\kappa^n}{n!} \,
[\, \csf_Q^n \,,B_1\, ,\, B_2 \,, \ldots, B_m \,] \,.
\label{primed-products}
\end{equation}
The operator $Q'$  satisfies
\begin{equation}
\langle Q' A \,, B \rangle = (-1)^A  \, \langle A \,, Q' B \rangle \,.
\end{equation}
It is also natural to introduce
graded-commutative shifted multilinear forms:
\begin{equation}
\bigl\{  B_1\, ,\, B_2 \,, \ldots, B_n\bigr\}' = \bigl\langle  B_1,
[\, B_2\, , \ldots\,, B_n\,]' \bigr\rangle \,.
\end{equation}

As shown in equation (4.117) of~\cite{Zwiebach:1992ie},
\begin{equation}
\label{qprimesquared}
Q' (Q' A) = - \kappa \, [\, \mathcal{F}(\csf_Q) \,, A\, ]' \,.
\end{equation}
In subsection \ref{3.2} we  define the string field $\csf_Q$
explicitly and show that it satisfies the equation of
motion $\mathcal{F}(\csf_Q) =0$.
As a result, $Q'$ squares to zero. Moreover,
$Q'$ and the new products also satisfy
the main identity that the original BRST operator and products do.
For example, $Q'$ is a derivation of the primed product with two inputs:
\begin{equation}
Q' \, [\, B_1 ,\, B_2 \,]'
= {}- [\, Q' B_1 ,\, B_2 \,]' -(-1)^{B_1} [\, B_1 ,\, Q' B_2 \,]'  .
\label{Q'-derivation}
\end{equation}
Just as $Q$ fails to be a derivation of $[\,\cdot\,, \cdot\,, \cdot\,]$,
$Q'$ fails to be a derivation of
$[\,\cdot\,, \cdot\,, \cdot\,]'$:
\begin{eqnarray}
\label{failjacobi}
0 &=& \hskip-5pt Q' \, [\, B_1,\, B_2,\, B_3 \,]'
+ [\, Q'B_1,\, B_2,\, B_3 \,]'
+ (-1)^{B_1} \, [\, B_1,\, Q'B_2,\, B_3 \,]'
+ (-1)^{B_1+B_2} \, [\, B_1,\, B_2,\, Q'B_3 \,]'
\nonumber\\[1ex] &&\hskip-8pt
{}+ (-1)^{B_1} \, [\, B_1,\, [\, B_2,\, B_3 \,]' \,]'
+ (-1)^{B_2 (1+ B_1)} \, [\, B_2,\, [\, B_1,\, B_3 \,]' \,]'
+ \, [\,  [\, B_1,\, B_2 \,]' \,,B_3\,]' \,.
\end{eqnarray}

Derivations of all of the string products
play an important role in our construction of
heterotic string field theory.
An operation $X$ is a derivation if $X$ satisfies
\begin{eqnarray}
X \, [\, B_1 ,\, B_2 , \ldots ,\, B_m \,]
&=& (-1)^X [\, X B_1 ,\, B_2 , \ldots ,\, B_m \,]
+ (-1)^{X (1+B_1)} [\, B_1 ,\, X B_2 , \ldots ,\, B_m \,]
\nonumber \\
&& {}+ \cdots
+ (-1)^{X (1+B_1+B_2+ \cdots +B_{m-1})}
[\, B_1 ,\, B_2 , \ldots ,\, X B_m \,]
\label{derivation-definition}
\end{eqnarray}
for $m \ge 2$.
In this paper we only consider derivations
that further satisfy $X Q = (-1)^X Q X$.
In fact, the graded-commutativity of $X$ with $Q$
is the statement that
(\ref{derivation-definition}) holds for $m=1$
because the one-input product is defined by
$[\, B \,] \equiv Q B$~\cite{Zwiebach:1992ie}.
While $Q$ and $X$ commute (or anticommute),
$Q'$ and $X$ do not.
It is easy to show that for arbitrary string field $B$
\begin{equation}
Q' (X B) - (-1)^X  \, X \, (Q' B)
= -\kappa \, [\, X \csf_Q ,\, B \,]' \,.
\label{Q'-X-commutator-closed}
\end{equation}

\subsection{Pure-gauge closed string fields}\label{3.2}

Let us consider a `large' pure-gauge closed string field $\G (V)$
associated with a finite gauge parameter~$V$.
Such pure-gauge field can be built
by successive infinitesimal gauge transformations
but, as in any non-Abelian theory, the result depends on the path
connecting $0$ and $V$.
The pure-gauge open string field $e^{-\Phi} (Q e^{\Phi})$
corresponds to choosing a straight line connecting $0$ and $\Phi$.
For closed strings we also choose a straight line
connecting $0$ and $V$
and parametrize the path as $\tau V$ with $0 \le \tau \le 1$.
Intermediate pure-gauge fields are given by $\G (\tau V)$,
and the final pure-gauge string field $\G (V)$
corresponds to $\G (\tau V)$ at $\tau = 1$.
This string field is defined by the following two conditions.
First, $\G (\tau V)$ vanishes at $\tau=0$,
and second,  $\G (\tau V + d \tau V)$ and $\G (\tau V)$
differ by a gauge transformation with gauge parameter $d \tau V$:
\begin{equation}
\G (\tau V + d \tau V) - \G (\tau V)
= Q (d \tau V) + \sum_{n=1}^\infty \frac{\kappa^n}{n!} \,
[\, \G (\tau V)^n ,\, d \tau V \,] + \mathcal{O} (d \tau^2) \,.
\end{equation}
Therefore, $\G (\tau V)$ satisfies the differential equation
\begin{equation}
\partial_\tau \G (\tau V) = Q V
+ \sum_{n=1}^\infty \frac{\kappa^n}{n!} \,
[\, \G (\tau V)^n ,\, V \,]
\equiv Q'_\G V \,,
\label{G-definition}
\end{equation}
with the initial condition $\G (0) = 0$.

It is in fact straightforward
to solve (\ref{G-definition})
in an expansion in powers of $\kappa$.
Let us expand the equation and $\G$ as
\begin{equation}
\partial_\tau \G = QV
+ \kappa \, [\, \G ,\, V \,]
+ \frac{\kappa^2}{2} \, [\, \G ,\, \G ,\, V \,]
+ \mathcal{O} (\kappa^3) \,,
\end{equation}
and
\begin{equation}
\G = \G^{(0)} + \kappa \, \G^{(1)} + \kappa^2 \, \G^{(2)}
+ \mathcal{O} (\kappa^3) \,.
\end{equation}
The initial condition gives $\G^{(n)}=0$ at $\tau=0$ for all $n \ge 0$.
The equation at $\mathcal{O} (\kappa^0)$ is given by
$\partial_\tau \G^{(0)} = QV$ and is solved by
$\G^{(0)} = \tau \, QV$.
The equation at $\mathcal{O} (\kappa)$ is
$\partial_\tau \G^{(1)} = [\, \G^{(0)} ,\, V \,]
= [\, \tau \, QV ,\, V \,]$
and is solved by
$\G^{(1)} = \frac{\tau^2}{2} \, [\, QV ,\, V \,] \,.$
It is straightforward to proceed to higher orders in this way.
After setting $\tau =1$, the first few terms of $\G (V)$ are given  by
\begin{equation}
\label{csfpgaugeformula}
\G (V) = QV + \frac{\kappa}{2} \, [\, V,\, QV \,]
+ \frac{\kappa^2}{3!} \, [\, V,\, QV,\, QV \,]
+ \frac{\kappa^2}{3!} \, [\, V,\, [\, V,\, QV \,] \,]
+ \mathcal{O} (\kappa^3) \,.
\end{equation}

Large gauge transformations in closed string field theory were considered
long ago by Schubert \cite{Schubert:1991en}.  The class of terms
that appear are all those constructed with $V$ and $QV$ that are consistent
with ghost number~\cite{Okawa:2004ii}, and  simple rules to determine
the numerical coefficient of a given term were stated
in~\cite{Schubert:1991en}.
Our result (\ref{csfpgaugeformula})
is in agreement with the rules in \cite{Schubert:1991en}.

Since $\G (V)$ is a pure gauge, we expect it
to satisfy the closed string field theory equation of motion.
This is readily proven by evaluating
$\partial_\tau \mathcal{F} (\G (\tau V))$.
Using (\ref{G-definition}) and (\ref{qprimesquared}) we find
\begin{eqnarray}
\partial_\tau \mathcal{F} (\G (\tau V))
&=& \partial_\tau \, \Bigl(  Q \G (\tau V) +
\sum_{n=2}^\infty \frac{\kappa^{n-1}}{n!} \, [\, \G (\tau V)^n
\, ] \Bigr)
\nonumber \\
&=& Q ( \partial_\tau \G (\tau V))
+\sum_{n=1}^\infty \frac{\kappa^n}{n!} \, [\, \G (\tau V)^n ,\,
\partial_\tau \G (\tau V) \,]
\nonumber \\
&=& Q'_\G ( \partial_\tau \G (\tau V) )
= Q'_\G ( Q'_\G V )
= - \kappa \, [ \mathcal{F}( \G (\tau V) ),\, V \,]'_\G \,,
\end{eqnarray}
where a primed product with subscript $\G$ is defined
by (\ref{primed-products}) with $\csf_Q$ replaced by $\G (\tau V)$.
We therefore have the following differential equation
for $\mathcal{F} (\G (\tau V))$:
\begin{equation}
\partial_\tau \mathcal{F} (\G (\tau V))
= \kappa \, [\, V,\, \mathcal{F}( \G (\tau V) ) \,]'_\G \,.
\end{equation}
Since $\G (\tau V)$ vanishes at $\tau=0$,
$\mathcal{F} (\G (0)) = 0$.
The above first-order differential equation is solved
by $\mathcal{F} (\G (\tau V))=0$ for any $\tau$
so, by uniqueness, this is the solution
with the given initial condition.
After setting $\tau=1$,
we thus obtain that
\begin{equation}
\label{csfeom}
\mathcal{F}(\G (V)) = Q \G (V)
+ \sum_{n=2}^\infty \frac{\kappa^{n-1}}{n!} \,
[\, \G (V)^n \,] = 0 \,.
\end{equation}

In open superstring field theory
$A_Q$ is a pure-gauge open string field
for each value of $t$.  To define an analogous
closed string field we proceed as follows.
For any given string field $V$ we introduce
a $t$-dependent string field $V(t)$, with $t\in [0,1]$, that connects
the zero string field
$0=V(t=0)$ and $V=V(t=1)$. We  now define $\csf_Q$ by
\begin{equation}
\csf_Q \equiv \G (V(t)) \,.
\label{E-definition}
\end{equation}
The string field $\csf_Q$ is a pure-gauge closed string field
for each value of $t$.  Since $\csf_Q$ satisfies the string field
equation of motion,
the operator
\begin{equation}
Q'\equiv Q'_{\csf_Q} \,,
\end{equation}
and the primed products satisfy
the familiar identities of closed string field theory.
In particular, $Q'{}^2$ vanishes.

\subsection{Derivations and associated string fields}

Let us next construct $\csf_\dt$, $\csf_\delta$, and $\csf_\eta$
satisfying (\ref{E_partial-equation}), (\ref{E_delta-equation}),
and (\ref{E_eta-equation}). As in open superstring field theory,
the existence of $\csf_\delta$ is guaranteed because
   $\csf_Q$ is a pure-gauge string field and therefore
any infinitesimal
variation of $\csf_Q$ induced by a variation  $\delta V(t)$
can be written as a gauge transformation
$Q' \csf_\delta$ for some $\csf_\delta$.  The existence of
$\csf_\dt$ and $\csf_\eta$ also follows since derivations
can be viewed as formal variations.

We claim that the problem of explicit construction of these string fields
reduces to
that of constructing a string field $\HH (V, XV)$ which satisfies
\begin{equation}
X \G (V) = (-1)^X Q'_\G \HH (V, XV) \,,
\label{H-equation}
\end{equation}
where $X$ is any derivation of all of the closed string products
and satisfies $[\, X ,\, Q \,\} =0$.
We view equation (\ref{H-equation})  as a relation between maps
$V \to G(V)$ and $(V, XV) \to H(V,XV)$ that holds for arbitrary
$V$ on which the action of $X$ is well defined.

Let us now demonstrate that (\ref{H-equation}) provides a construction
of $\csf_\dt$, $\csf_\delta$, and $\csf_\eta$.   First note that equation
(\ref{H-equation}) holds for any fixed
$t$ if we replace each $V$ by $V(t)$.
Since $\csf_Q = \G (V(t))$ and we write $Q' = Q_{\csf_Q}'$,
    we have
\begin{equation}
X \csf_Q = (-1)^X \,Q' \,\HH \bigl(V(t), XV(t)\bigr) \,.
\label{H1-equation}
\end{equation}
Since $\delta$ and $\eta$ are derivations that
act with fixed $t$,  the above equation clearly holds for
$X = \delta$ and $X=\eta$.  We thus find that
$\csf_\delta$ and $\csf_\eta$ satisfying (\ref{E_delta-equation})
and (\ref{E_eta-equation}) are given by
$\csf_\delta = \HH (V(t), \delta V(t))$
and $\csf_\eta = \HH (V(t), \eta V(t))$.
In fact, equation (\ref{H1-equation}) also holds for $X = \partial_t$.
By construction,
$\csf_Q$ depends on
$t$ only through
$V(t)$ so $\partial_t \csf_Q = \delta \csf_Q$
with $\delta V(t) = \partial_t V(t)$.
Since $\delta \csf_Q = Q' \csf_\delta = Q' \HH (V(t), \delta V(t))$,
$\csf_\dt$ satisfying (\ref{E_partial-equation})
is given by $\HH (V(t), \partial_t V(t))$.  In  summary,
\begin{eqnarray}
\label{edf1}
\csf_\dt &=& \HH (V(t), \partial_t V(t)) \,,
\\
\label{edf2}
\csf_\delta &=& \HH (V(t), \delta V(t)) \,,
\\
\label{edf3}
\csf_\eta &=& \HH (V(t), \eta V(t)) \,.
\end{eqnarray}
Equation (\ref{H1-equation}) and the equations
above are summarized as
\begin{equation}
X\csf_Q  = (-1)^X \, Q' \csf_X \,,  \quad \hbox{with}
    \quad \Psi_X = H( V(t), X V(t)) \,,
\end{equation}
for $X = \partial_t, \delta$, and $\eta$.

In open string field theory,
the string field $A_X$ that satisfies $X A_Q = (-1)^X Q' A_X$
is obtained by replacing $Q$ by $X$  in the expression for $A_Q$.
In the case of closed string field theory, however,
the construction of $\HH (V, XV)$ or, equivalently, $\csf_X$
is more complicated.
As in the case of $\G (V)$,
we construct $\HH (V, XV)$
as the solution to a differential equation for $\HH (V, XV; \tau)$,
evaluated at $\tau =1$.
To derive the requisite differential equation we define the auxiliary
string field
$\mathcal{H}(\tau)$ inspired by the
$\tau$-dependent version of  (\ref{H-equation}):
\begin{equation}
\mathcal{H} (\tau)\equiv
Q'_\G \HH (V, XV; \tau) -(-1)^X X \G (\tau V)\,.
\end{equation}
Equation (\ref{H-equation}), which we want to prove,
is equivalent to $\mathcal{H}(1)=0$.
We compute $\partial_\tau \mathcal{H}$
using
(\ref{Q'-X-commutator-closed})
and (\ref{G-definition}).  The result is
\begin{eqnarray}
\partial_\tau \mathcal{H}
= \kappa \, [\, V,\, \mathcal{H} \,]'_\G
+ Q'_\G \,\Bigl(\, \partial_\tau \HH (V, XV; \tau)
- X V - \kappa \, [\, V ,\, \HH (V, XV; \tau) \,]'_\G \,\Bigr) \,.
\end{eqnarray}
We will demand that the expression in parentheses on the right-hand
side vanishes:
\begin{equation}
\label{hvxveqn}
\partial_\tau \HH (V, XV; \tau)
= X V + \kappa \, [\, V ,\, \HH (V, XV; \tau) \,]'_\G \,.
\end{equation}
With  this differential equation for $\HH (V, XV; \tau)$,
the differential equation for $\mathcal{H}$ becomes
\begin{equation}
\partial_\tau \mathcal{H}
= \kappa \, [\, V,\, \mathcal{H} \,]'_\G \,.
\end{equation}
With initial condition $\mathcal{H}(\tau=0) = 0$,
this differential equation ensures that $\mathcal{H}$ vanishes
for any~$\tau$, and  (\ref{H-equation}) is proven.
We must therefore solve (\ref{hvxveqn}) with initial conditions that
ensure that
$\mathcal{H}(\tau=0) = 0$, namely,
\begin{equation}
\label{auxcondde}
Q'_\G \HH (V, XV; \tau) = (-1)^X X \G (\tau V)\qquad  \hbox{at}\quad
\tau = 0\,.
\end{equation}
Note that $X \G (\tau V)$ vanishes at $\tau = 0$ because $X$ does not
act on the
auxiliary parameter $\tau$ and $\G (\tau V)$ vanishes for $\tau=0$
for any finite $V$.
We can then require
\begin{equation}
\label{dflkoiekjb}
\HH (V, XV; \tau=0) = 0\,,
\end{equation} and (\ref{auxcondde})
holds with both terms equal to zero.  As in
the case of the differential equation for $\G (\tau V)$, it is
straightforward to solve
(\ref{hvxveqn}) in powers of $\kappa$.  With the initial  condition
(\ref{dflkoiekjb})  we find
that the first few terms in $\HH (V, XV)$ are given by
\begin{equation}
\HH (V, XV) = XV
+ \frac{\kappa}{2} \, [\, V,\, XV \,]
+ \frac{\kappa^2}{3} \, [\, V,\, QV,\, XV \,]
+ \frac{\kappa^2}{3!} \, [\, V,\, [\, V,\, XV \,] \,]
+ \mathcal{O} (\kappa^3) \,.
\label{perturbative-H}
\end{equation}
This result can be used to write  explicit expressions
for $\csf_\dt$, $\csf_\delta$, and $\csf_\eta$ using (\ref{edf1}),
(\ref{edf2}), and (\ref{edf3}).

\subsection{Heterotic string field theory action}

Having constructed $\csf_Q$, $\csf_\dt$, $\csf_\delta$,
and $\csf_\eta$ satisfying (\ref{E_partial-equation}),
(\ref{E_delta-equation}), and (\ref{E_eta-equation}),
we now show that the action
(\ref{hsfta})
has the requisite gauge invariances.
The proof is completely parallel
to the one for open superstring field theory.

In the computation of $\delta S$
we will use the identity:
\begin{equation}
Q' \, (\, \partial_t \csf_\delta - \delta \csf_\dt
- \kappa \, [\, \csf_\dt ,\, \csf_\delta \,]' \,) = 0 \,.
\end{equation}
This identity, which is the closed string analog of
(\ref{theextraone}),
follows from (\ref{E_partial-equation})
and (\ref{E_delta-equation})
with the help of (\ref{Q'-X-commutator-closed}):
\begin{eqnarray}
0 &=& \partial_t ( Q' \csf_\delta ) - \delta ( Q' \csf_\dt )
\nonumber \\
&=& Q'(\partial_t \csf_\delta)
+ \kappa \, [\, \partial_t \csf_Q ,\, \csf_\delta \,]'
- Q'(\delta \csf_\dt)
- \kappa \, [\, \delta \csf_Q ,\, \csf_\dt \,]'
\nonumber \\
&=& Q'(\, \partial_t \csf_\delta - \delta \csf_\dt
- \kappa \, [\, \csf_\dt ,\, \csf_\delta \,]' \,) \,.
\end{eqnarray}
Let us compute the variation of the integrand of the action:
\begin{equation}
\delta \, \langle \, \eta \csf_\dt ,\, \csf_Q \, \rangle
= \langle \, \eta \, \delta \csf_\dt ,\, \csf_Q \, \rangle
+ \langle \, \eta \csf_\dt ,\, \delta \csf_Q \, \rangle \,.
\end{equation}
The first term on the right-hand side is evaluated as follows:
\begin{eqnarray}
\langle \, \eta \, \delta \csf_\dt ,\, \csf_Q \, \rangle
&=& - \langle \, \delta \csf_\dt ,\, \eta \csf_Q \, \rangle
= \langle \, \delta \csf_\dt ,\, Q' \csf_\eta \, \rangle
= - \langle \, Q' \delta \csf_\dt ,\, \csf_\eta \, \rangle
\nonumber \\
&=& - \langle \, Q' (\, \partial_t \csf_\delta
+ \kappa \, [\, \csf_\delta ,\, \csf_\dt \,]' \,) ,\,
\csf_\eta \, \rangle
\nonumber \\
&=& - \langle \, \partial_t \csf_\delta
+ \kappa \, [\, \csf_\delta ,\, \csf_\dt \,]' ,\,
\eta \csf_Q \, \rangle
\nonumber \\
&=& \langle \, \eta \, \partial_t \csf_\delta ,\, \csf_Q \, \rangle
- \kappa \, \langle \, \eta \csf_Q ,\,
[\, \csf_\delta ,\, \csf_\dt \,]' \, \rangle
\nonumber \\
&=& \langle \, \eta \, \partial_t \csf_\delta ,\, \csf_Q \, \rangle
- \kappa \, \{\, \eta \csf_Q ,\,
\csf_\delta ,\, \csf_\dt \,\}' \,.
\end{eqnarray}
The second term on the right-hand side is evaluated as follows:
\begin{eqnarray}
\langle \, \eta \csf_\dt ,\, \delta \csf_Q \, \rangle
&=& \langle \, \eta \csf_\dt ,\, Q' \csf_\delta \, \rangle
= \langle \, Q' (\eta \csf_\dt) ,\, \csf_\delta \, \rangle
\nonumber \\
&=& - \langle \, \eta \, (Q' \csf_\dt) ,\, \csf_\delta \, \rangle
- \kappa \, \langle \,
[\, \eta \csf_Q ,\, \csf_\dt \,]' ,\, \csf_\delta \, \rangle
\nonumber \\
&=& - \langle \, \eta \, \partial_t \csf_Q ,\,
\csf_\delta \, \rangle
- \kappa \, \langle \, [\, \eta \csf_Q ,\, \csf_\dt \,]' ,\,
\csf_\delta \, \rangle
\nonumber \\
&=& \langle \, \eta \csf_\delta ,\, \partial_t \csf_Q \, \rangle
- \kappa \, \langle \, \csf_\delta ,\,
[\, \eta \csf_Q ,\, \csf_\dt \,]' \, \rangle
\nonumber \\
&=& \langle \, \eta \csf_\delta ,\, \partial_t \csf_Q \, \rangle
+ \kappa \, \{\, \eta \csf_Q ,\, \csf_\delta ,\, \csf_\dt \,\}' \,.
\end{eqnarray}
Therefore, the variation of the integrand of the action
is given by
\begin{equation}
\delta \, \langle \, \eta \csf_\dt ,\, \csf_Q \, \rangle
= \partial_t \, \langle \, \eta \csf_\delta ,\, \csf_Q \,
\rangle \,.
\end{equation}
Since this variation
is a total $t$ derivative, the variation of the action
is given by
\begin{equation}
\delta S = \frac{2}{\alpha'} \, \langle \,
\eta \, \overline{\csf}_\delta ,\, \overline{\csf}_Q \, \rangle
= - \frac{2}{\alpha'} \, \langle \,
\overline{\csf}_\delta ,\, \eta \, \overline{\csf}_Q \, \rangle \,,
\label{delta-S-1}
\end{equation}
where the bar, as before,  denotes evaluation at $t=1$.
Using (\ref{E_eta-equation}),
the variation of the action can also be written as
\begin{equation}
\delta S = \frac{2}{\alpha'} \,
\langle \, \overline{\csf}_\delta ,\,
\overline{Q'} \, \overline{\csf}_\eta \, \rangle
= - \frac{2}{\alpha'} \, \langle \,
\overline{Q'} \, \overline{\csf}_\delta ,\,
\overline{\csf}_\eta \, \rangle \,.
\label{delta-S-2}
\end{equation}
Recall
that $\csf_\delta$ depends on $t$ only through $V (t)$
and $\delta V(t)$.
The first few terms of ${\csf}_\delta$
as a series in powers of $\kappa$ are
\begin{equation}
{\csf}_\delta = \delta V(t)
+ \frac{\kappa}{2} \, [\, V(t),\, \delta V(t) \,]
+ \frac{\kappa^2}{3} \, [\, V(t),\, QV(t),\, \delta V(t) \,]
+ \frac{\kappa^2}{3!} \, [\, V(t),\, [\, V(t),\, \delta V(t) \,] \,]
+ \mathcal{O} (\kappa^3) \,.
\label{perturbative-Psi_delta}
\end{equation}
It is straightforward to verify that
$\csf_\delta$ is an invertible function of $\delta V(t)$,
at least in the expansion in powers of $\kappa$.
One writes
$\delta V(t) = v^{(0)} + \kappa \, v^{(1)}
+ \kappa^2 \, v^{(2)} + \ldots$ \,,
substitutes in (\ref{perturbative-Psi_delta}),
and solves recursively for $v^{(0)}, v^{(1)}, v^{(2)}, \ldots$ \,,
without encountering obstructions.
The first few terms of the resulting expression are
\begin{equation}
\label{invertedeltaldeltav}
\delta V(t) =  {\csf}_\delta
+ \frac{\kappa}{2} \, [\, {\csf}_\delta,\, V(t) \,]
+ \frac{\kappa^2}{3} \, [\, {\csf}_\delta,\, QV(t),\, V(t) \,]
+ \frac{\kappa^2}{12} \,
[\, [\, {\csf}_\delta,\, V(t) \,] \,,\, V(t) \,]
+ \mathcal{O} (\kappa^3) \,.
\end{equation}
Since the relation between ${\csf}_\delta$ and $\delta V(t)$
is invertible,
arbitrary $\delta V$ is equivalent
to arbitrary $\overline\csf_\delta$.
We therefore conclude that (\ref{delta-S-1}) gives
the following equation of motion:
\begin{equation}
\eta \, \overline{\csf}_Q = 0 \,.
\end{equation}
The equation of motion can also be written as
$\overline{Q'} \, \overline{\csf}_\eta = 0$,
as is manifest from the expression (\ref{delta-S-2}).

Let us now consider the gauge invariances of the action.
The action is manifestly invariant
under a  variation
for which $\overline{\csf}_\delta
= \eta \, \Omega \,$:
\begin{equation}
\label{varetatract}
\delta_\Omega S = \frac{2}{\alpha'} \,
\langle \, \eta \, (\eta \Omega) \,,\,
\overline{\csf}_Q \, \rangle = 0 \,.
\end{equation}
We therefore have a gauge transformation
\begin{equation}
\csf_\delta = \eta\, \Omega(t) \,,
\label{Omega-gauge-transformation}
\end{equation}
where $\Omega(0)=0$ and $\Omega(1)=\Omega$.
The formula (\ref{invertedeltaldeltav})
can be used to find, if desired,
an explicit expression for the gauge variation $\delta_\Omega V(t)$
associated with (\ref{Omega-gauge-transformation}).
We immediately find that $\delta_\Omega V= \delta_\Omega V(t=1)$
is given by
\begin{equation}
\label{etagaugetrexpl}
\delta_\Omega \hskip1pt V
= \eta\hskip1pt\Omega
+ \frac{\kappa}{2} \, [\, \eta\hskip1pt\Omega\hskip1pt,\, V \,]
+  \,
\frac{\kappa^2}{3} \, [\, \eta\hskip1pt\Omega\hskip1pt,\, QV,\, V \,]
+ \frac{\kappa^2}{12} \, [\, [\, \eta\hskip1pt\Omega\hskip1pt,\, V \,] \,,
V \,] \,+ \mathcal{O}(\kappa^3) \,.
\end{equation}
By construction, the gauge parameter $\Omega$ always appears in
the form $\eta \Omega$. This was also the requirement used in
\cite{Okawa:2004ii} to fix the ambiguity from
field redefinition of the gauge parameter.
Therefore $\delta_\Omega V$ in (\ref{etagaugetrexpl})
should coincide with that in~\cite{Okawa:2004ii}.
We in fact found that this is the case
up to the order computed in~\cite{Okawa:2004ii}.

The action is also invariant
under a variation for which
$\overline{\csf}_\delta = \overline{Q'} \,\Lambda'$:
\begin{equation}
\label{varqgaugetract}
\delta_{\Lambda'} S = - \frac{2}{\alpha'} \, \langle \,
\overline{Q'} \, (\overline{Q'} \,\Lambda') \,,\,
\overline{\csf}_\eta \, \rangle = 0 \,.
\end{equation}
We therefore have the gauge transformation
\begin{equation}
{\csf}_\delta = {Q'}\Lambda'(t) \,,
\end{equation}
where $\Lambda'(0)=0$ and $\Lambda'(1)=\Lambda'$.
A formula for the associated gauge variation $\delta_{\Lambda'} V$
can be obtained  using (\ref{invertedeltaldeltav}) with
$\overline{\csf}_\delta = \overline{Q'}\,\Lambda'$.
When we expand this formula in powers of $\kappa$,
$\Lambda'$ sometimes appears in the form $Q\Lambda'$,
but sometimes it appears without $Q$.
The gauge transformation $\delta_\Lambda V$ in \cite{Okawa:2004ii}
was fixed by the requirement that the gauge parameter $\Lambda$ always
appear in the form $Q \Lambda$.
The transformations do not coincide exactly,
but they should be related by redefinition
of the gauge parameters. We in fact find
\begin{equation}
\Lambda' = \Lambda + \kappa\, [\,V\,, \Lambda\,] + {\kappa^2\over 2}\,
[\,V\,, QV\,, \Lambda\,]
+ {\kappa^2\over 2}\,  [\,V, [\,V, \Lambda\,]\,] +
\mathcal{O} (\kappa^3).
\end{equation}

\subsection{Reparameterization invariance and linear homotopy}

We have shown that the heterotic string field theory action
(\ref{hsfta}) is invariant under the gauge transformations
$\delta_{\Lambda'} V$ and $\delta_\Omega V$.
Moreover, the action is independent of the parametrized path $V(t)$
in field space that joins $V(0)=0$ to $V(1) = V$.
We established this fact
by showing that the variation of the action depends only on
the values of the variation and fields at $t=1$.
To write the string field theory action we can
therefore choose any path $V(t)$ from $0$ to $V$.
The path independence implies that
the action is invariant under reparametrization of any chosen path.
While the independence of the parametrized path is not manifest
in the action, this reparametrization invariance is manifest.
We first note that $V(t)$ is scalar
under the reparametrizations: $V(t) = V'(t')$.
Moreover,
$\csf_Q$ and $\csf_\dt$ transform~as
\begin{equation}
\label{partransfdod}
\csf_Q (V(t)) = \csf_Q' (V'(t')) \,, \qquad
\csf_\dt (V(t)) = \frac{dt'}{dt} \, \csf'_\dt (V'(t')) \,.
\end{equation}
The field $\csf_Q$ is a scalar
because it depends on $t$ only through $V(t)$.
The transformation of $\csf_\dt$ follows from
(\ref{E_partial-equation}) and the fact that
$Q'$ is a scalar.
The two equations in (\ref{partransfdod}) imply
the reparameterization
invariance of the action (\ref{hsfta}).

The action takes a simple form when we choose a straight path from
$0$ to $V$ and
we parameterize it linearly:
$V(t) = t V$.
In this case $\csf_Q = \G (t V)$ from
the definition (\ref{E-definition}) and
$\partial_t \csf_Q =\partial_t G(tV) =  Q' V$
using (\ref{G-definition}).
Comparing with (\ref{E_partial-equation}) we conclude that
$\csf_\dt = V$, and the action (\ref{hsfta}) becomes
\begin{equation}
\label{therkljdflk49}
S = \frac{2}{\alpha'} \, \int_0^1 dt \,
\langle \, \eta V ,\, G(tV)  \, \rangle \,.
\end{equation}
This form of the action is
very useful.
The coefficient of any term in $\csf_Q = \G (t V)$
can be easily obtained from Schubert's rules \cite{Schubert:1991en},
and the integral over $t$ is trivial.
Using the expression for $G(tV)$ as a series in
$\kappa$, the first few terms of the action are given by
\begin{equation}
S = \frac{2}{\alpha'}\int_{0}^{1} dt \,
\bigl\langle \, \eta V,\,  t QV  + {\kappa\over 2} t^2 [V, QV]
+ \ldots  \,
\bigr\rangle  = \frac{2}{\alpha'}\Bigl( {1\over 2}\,
\langle \eta V\,, QV\rangle + {\kappa\over 3!} \,
\langle \,\eta V, [V, QV]\,\rangle + \ldots\Bigr) \,.
\end{equation}
These terms agree with those in the action given in~\cite{Okawa:2004ii}.
In fact, one can easily check the agreement for all the
terms that were determined in~\cite{Okawa:2004ii}.

\sectiono{Heterotic string action in standard WZW form}

We have seen that the familiar WZW bosonic action (\ref{wzwconventional}),
which involves a two-dimensional and a three-dimensional
term, can be written in the less familiar compact form (\ref{newf}).
A similar fact is true for open superstrings: the original WZW action
(\ref{wzwopenstring})
can be written as in (\ref{ossfta}).
We would like to know if the heterotic string action
\begin{equation}
\label{newiondesired}
S =  \frac{2}{\alpha'} \int_{0}^{1} dt \,
\langle \, \eta \csf_\dt ,\, \csf_Q \, \rangle \,.
\end{equation}
can be transformed into the more familiar WZW form.
  We find that such
a familiar form is obtained when $V(t) = tV$, but
there is an additional term for general paths $V(t)$.

We begin by a straightforward attempt to transform the action
(\ref{newiondesired}). Our experience in performing
a similar transformation for open superstrings indicates that
in addition to (\ref{E_partial-equation}) and
(\ref{E_eta-equation}), we must use the closed string analog of
equation (\ref{dlkgoiiejnf}), which states that the
field strength $F_{\eta \dt}$ vanishes.   Nevertheless, for
closed strings we find that the following combination $E_{\eta\dt}$
of operators
does not vanish:
\begin{equation}
\label{xy5}
E_{\eta\dt}\equiv \eta  \csf_\dt- \partial_t \csf_\eta + \kappa \,
[\csf_\eta\, , \csf_\dt\,]' \not=0 \,.
\end{equation}
Indeed, $\csf_\dt$ and $\csf_\eta$ have already been defined, and
a computation shows that $E_{\eta \dt}$ does
not vanish for arbitrary $V(t)$.
We proceed, anyway, by writing
\begin{eqnarray}
S &=& {1\over \alpha'} \int_{0}^{1} dt \,
\langle \, \eta \csf_\dt ,\, \csf_Q \, \rangle
- {1\over \alpha'} \int_{0}^{1} dt \,
\langle \,  \csf_\dt ,\, \eta \csf_Q \, \rangle \nonumber  \\
&=& {1\over \alpha'} \int_{0}^{1} dt \,
\bigl\langle \, \partial_t \csf_\eta
-\kappa \, [\, \csf_\eta ,\, \csf_\dt \,]'
,\, \csf_Q \, \bigr\rangle
+ {1\over \alpha'} \int_{0}^{1} dt \,
\langle \, E_{\eta\dt} ,\, \csf_Q \, \rangle
+ {1\over \alpha'} \int_{0}^{1} dt \,
\langle \,  \csf_\dt ,\, Q'\csf_\eta \, \rangle\,.
\end{eqnarray}
A little reorganization gives
\begin{equation}
S = {1\over \alpha'} \int_{0}^{1} dt \,
\bigl( \bigl\langle \, \partial_t \csf_\eta ,\, \csf_Q \, \bigr\rangle
+\langle \,  \csf_\eta ,\, Q'\csf_\dt\rangle\bigr)
+ {\kappa \over \alpha'} \int_0^1 dt \,
\langle
\,\csf_Q ,\, [\, \csf_\eta\,,\,\csf_\dt\,{]\hskip 0.8pt}' \,
\rangle\, +
{1\over \alpha'} \int_{0}^{1} dt \,
\langle \, E_{\eta\dt} ,\, \csf_Q \, \rangle .
\end{equation}
Using (\ref{E_partial-equation}) and the  graded commutativity of
the multilinear forms, we have
\begin{equation}
S = {1\over \alpha'} \int_{0}^{1} dt \, \partial_t
\bigl\langle \, \csf_\eta ,\, \csf_Q \, \bigr\rangle
+ {\kappa \over \alpha'} \int_0^1 dt \,
\langle
\,\csf_\dt\,,\, [\, \csf_\eta\,,\,\csf_Q \,{]\hskip 0.8pt}' \,
\rangle\, +
{1\over \alpha'} \int_{0}^{1} dt \,
\langle \, E_{\eta\dt} ,\, \csf_Q \, \rangle .
\end{equation}
Integrating the first term we find
\begin{equation}
\label{finhetstractwzw}
S = {1\over \alpha'} \,\Bigl(\,
\langle \,\, \overline{\csf}_\eta ,\, \overline{\csf}_Q \,
\rangle + \, \kappa \int_0^1 dt \,
\langle
\,\csf_\dt \,,\, [\, \csf_\eta \,,\, \csf_Q \,{]\hskip 0.8pt}' \,
\rangle\,\, +
\int_{0}^{1} dt \,
\langle \, E_{\eta\dt} ,\, \csf_Q \, \rangle \Bigr)\, .
\end{equation}
The last term in the action is reparameterization invariant
and is necessary for the equivalence with (\ref{newiondesired}).
The above action is unusual but represents some kind of
generalized WZW form.  In ordinary WZW theory, the connections are
flat and all field strengths vanish.  The open string analog of
$E_{\eta\dt}$ is the vanishing field strength $F_{\eta\dt}$.

\medskip
We now prove that $E_{\eta\dt}$ vanishes when we choose a linear
path $V(t) = t V$.   We have shown  in the lines above equation
(\ref{therkljdflk49}) that in this case $\csf_\dt = V$.
So we would
like to show that
\begin{equation}
\label{xyuu}
\eta  V - \partial_t \csf_\eta + \kappa \,
[\, \csf_\eta\, , V\,]' =0 \,.
\end{equation}
Consider equation (\ref{hvxveqn}) with $\tau$ replaced by $t$:
\begin{equation}
\label{hvxveqntt}
\partial_t \HH (V, XV; t)
= X V + \kappa \, [\, V ,\, \HH (V, XV; t) \,]' \,.
\end{equation}
It is straightforward to show from (\ref{hvxveqn}) that the solution $H$
satisfies the rescaling property
\begin{equation}
H\bigl( V, XV;  \tau\bigr) =  H \,\Bigl(\,  {1\over a} V,  {1\over a}
XV; a\tau \,\Bigr)\,,
\end{equation}
for any constant $a$.
Using this rescaling, equation (\ref{hvxveqntt}) becomes
\begin{equation}
\label{hvxveq9ntt}
\partial_t \HH (tV, XtV; 1)
= X V + \kappa \, [\, V ,\, \HH (tV, XtV; 1) \,]' \,.
\end{equation}
Since $tV = V(t)$, and the $H$ functions are evaluated at $t=1$, we have
\begin{equation}
\label{hvxveq99ntt}
\partial_t  \csf_X
= X V + \kappa \, [\, V ,\, \csf_X \,]' \,.
\end{equation}
For $X= \eta$, this equation is equivalent to (\ref{xyuu}), which is
what we wanted to establish. As a result
\begin{equation}
\label{fisdftstractwzw}
S = {1\over \alpha'} \,\Bigl(\,
\langle \,\, \overline{\csf}_\eta ,\, \overline{\csf}_Q \,\, \rangle
+ \, \kappa \int_0^1 dt \,
\langle
\,\csf_\dt\,,\, [\, \csf_\eta\,,\, \csf_Q \,{]\hskip 0.8pt}' \,
\rangle\,\,\Bigr)\, ,  \qquad  V(t) = t V \,.
\end{equation}
This is our closest
analog of the familiar WZW form.  Note that the action
is formally cubic in the `connections'
$\csf_Q, \csf_\dt$, and $\csf_\eta$.  The shifted product
$[\,\,\cdot\,,\,\,\cdot ]\,'$
contains additional dependence on $\csf_Q$, but the above
expression and our earlier analysis show that it is
a natural ingredient.  Some of the nonpolynomiality of closed
string field theory appears in the construction of
$\csf_Q, \csf_\dt$, and $\csf_\eta$, and further nonpolynomiality
is subsumed in the shifted product.  In open string field theory, there
is no shifted product because the action is cubic.

\medskip
To gain some understanding of the above
complications, we discuss the definition
of field strengths for closed strings.
We explained right after equation (\ref{009ifgkjej})
that the action of the
shifted BRST operator $Q'$ in open superstrings is that of a 
covariant derivative.
We thus set
\begin{equation}
\nabla_Q  = Q' \,.
\end{equation}
In ordinary field theory the definition
$F_{\mu\nu} \equiv [\, \nabla_\mu\,, \nabla_\nu \,]$
means that $F_{\mu\nu}$ is read from the equation $F_{\mu\nu} \cdot A
= \nabla_\mu (
\nabla_\nu A) -
\nabla_\nu (
\nabla_\mu A)$, valid for arbitrary $A$. For closed strings  we use
the two-input product to write
\begin{equation}
[\, F_{QQ}\,, A\, ] '
\equiv  \{ \nabla_Q , \nabla_Q\} A  = 2 Q' (Q'A) \,.
\end{equation}
Using  (\ref{qprimesquared}) we find
\begin{equation}
[\, F_{QQ}\,, A\, ] ' = - 2 \kappa\, [\, \mathcal{F}(\csf_Q) , A
]'\qquad \to \quad
F_{QQ} = - 2\kappa \, \mathcal{F}(\csf_Q)\,.
\end{equation}
As desired, the field strength $F_{QQ}$ vanishes because
$\csf_Q$ satisfies the equation  of  motion.
For derivations $X$ of all string products (that commute or
anti-commute with $Q$) we introduce covariant derivatives~as
\begin{equation}
\nabla_X  B \equiv  X B - (-1)^{X} \kappa \,
[\, \csf_X\,, B\,]'\,.
\end{equation}
We can now try to  define the field strengths that involve $Q$ and one
derivation $X$ of all closed string products by
$[\, F_{QX} \,, A \,]' \equiv  [ \nabla_Q\,, \nabla_X \} A $.  A
quick evaluation
of the right hand side shows that the definition is consistent and
\begin{equation}
\label{fqxdef}
F_{QX}  = -\kappa \, ( X \csf_Q - (-1)^{X}  Q' \csf_X)  \,.
\end{equation}
As desired, the commutator of covariant derivatives has produced
the quantity that we know vanishes.
This construction furnishes vanishing field strengths
$F_{Q\eta},$ and $F_{Q\dt}$, and $F_{Q\delta}$.

So far, our development has proceeded as well as it did for
open strings. Consider now
two  derivations
$X$ and $Y$ which commute or anticommute.
We are naturally led to define a field strength $F_{XY}$ by
$[ F_{XY}\,, A \, ]'  \equiv [\, \nabla_X\,, \nabla_Y\} A$.  This 
time, however,
an evaluation of the right-hand side shows that it is {\em not} of
the form $[ \cdots \,, A\,]'$
because the Jacobi identity does not hold strictly
in closed string theory (see (\ref{failjacobi})).
We thus fail to define the field strength~$F_{XY}$.
It is nevertheless clear that a certain combination $E_{XY}$ (that we
cannot call a field strength) plays a significant role:
\begin{equation} E_{XY} \equiv   X\,
\csf_Y - (-1)^{XY}
\, Y
\, \csf_X - (-1)^X \, \kappa \, [\, \csf_X \,, \csf_Y \,]'\,.
\end{equation}
The heterotic string action would take the standard WZW form  if
$E_{\eta\dt}$ vanished for arbitrary paths.   A direct
computation using the formulae for $\csf_X$ and $\csf_Y$ makes it clear
that $E_{XY}$ does not
generally vanish:\footnote{It is a nontrivial question if one can
make $E_{XY}=0$ by choosing appropriate
representatives for $\csf_X$ and $\csf_Y$
(recall $\csf_X \sim \csf_X + Q' \Lambda_X$
and $\csf_Y \sim \csf_Y + Q' \Lambda_Y$).
}
\begin{equation}
\label{fdlkgfkjndf}
E_{XY} = -\,{1\over 3} \, \kappa^2  \,
Q'\,[ \,XV(t)\,, V(t)\,, YV(t) ]' + \mathcal{O}(\kappa^3) \,.
\end{equation}
One can use $F_{QX}=0$ and $F_{QY}=0$ to verify that
$Q' E_{XY} =0$.  Indeed, $Q' E_{\delta\dt}=0$ was used to prove
the gauge invariance of the heterotic string theory.
A general argument indicates that
$Q'$ has no cohomology\footnote{
In the large Hilbert space $\eta$ has trivial cohomology.
Similarly, the BRST operator $Q$
has trivial cohomology.
This follows from  $\{Q \,,\, R\} =1$,
where $R$ is the zero mode of $c \xi\partial\xi e^{-2\phi}$.
Any vertex operator $V$ satisfying $QV=0$ can be written as
$V= \{Q \,,\, R \} V = Q R V$,
so $Q$ has trivial cohomology.
We now use an expansion
in powers of $\kappa$ to show that $Q'$, which differs from $Q$
by terms with positive powers of $\kappa$,
also has trivial cohomology.
Consider a state  $V = \sum_{n=0}^\infty  \kappa^n V_n$.
Then $Q' V =0$ implies that $Q V_0=0$, and consequently
$V_0 = Q\Omega_0$.
Therefore $(V - Q' \Omega_0) = \sum_{n=1}^\infty  \kappa^n W_n$
for some $W_n$ where $n\geq 1$.
This time,  $Q' (V -Q' \Omega_0) =0$ implies
that $W_1 = Q\Omega_1$, using a similar construction.
Continuing this procedure forever, one finds
$V = Q' (\sum_{n=0}^\infty  \Omega_n)$ for some $\Omega_n$.
So $Q'$ has trivial cohomology in the large Hilbert space.}
so we expect
$E_{XY} = Q' \Lambda_{XY}$ for some $\Lambda_{XY}$.
It remains to be investigated if there exist satisfactory definitions
of field strengths in closed string theory.

\sectiono{Conclusions}

In this paper we proposed the following action
for the NS sector of heterotic string field theory:
\begin{equation}
\label{repeat}
S=\frac{2}{\alpha'} \int_0^1 dt \,
\langle \, \eta \csf_\dt, \csf_Q \, \rangle \,.
\end{equation}
Here $\csf_Q = G(V(t))$, where $G(V)= QV + \ldots$
is a pure-gauge solution
of the bosonic closed string field theory equation of motion.
The action uses an NS string field $V(t)$ in the large Hilbert space,
with $V(0)=0$ and $V(1) = V$.
Because of its topological properties, the above action
depends only on $ V$.
The string field $\csf_\dt$ satisfies
$\partial_t \csf_Q = Q' \csf_\dt$,
where $Q'$ is the BRST operator around the background~$\csf_Q$.

The equation of motion implied by the action (\ref{repeat})
is $\eta \,\overline{\csf}_Q =0$,
where the bar denotes evaluation at $t=1$.
The action is invariant under nonlinear gauge transformations
of the form $\delta V = Q\Lambda + \eta \Omega + \, \ldots$ where
the terms indicated by dots can be explicitly determined.
This equation of motion and gauge invariance at linearized level
reproduce the heterotic NS spectrum, and the gauge invariance
is nonlinearly extended without using picture-changing operators.
These two features convince us that (\ref{repeat}) is correct.
A particularly simple and useful form of the action is
obtained by choosing $V(t) = tV$:
\begin{equation}
S={2\over \alpha'}\int_0^1 dt\, \langle \eta V,  G(tV)\, \rangle
\quad \hbox{when}
\quad V(t) = tV\,.
\end{equation}

The action (\ref{repeat}) is written as a WZW model
in the sense that
(1) gauge transformations generated by $Q$ and $\eta$
correspond to holomorphic and antiholomorphic gauge transformations,
(2) the action uses pure-gauge fields,
(3) the action is written using a third dimension,
and (4) the action takes a form which has an analog
in ordinary WZW theory.
Since closed string products have properties
altogether different from those of the open string star product,
some features of our heterotic action are new.
First, our action can be written in the standard WZW form
only when the string field $V(t)$
depends on the third dimension $t$ of the WZW model as
$V(t)=tV$ (see (\ref{fisdftstractwzw})).
Second, the language of flat connections and vanishing field strengths,
which works well in open superstring field theory (and ordinary
WZW theory), does not seem totally appropriate for heterotic strings.
Since the lowest closed string product $[\,\cdot\,, \, \cdot \,]$
does not satisfy the Jacobi identity,
it does not seem possible to define field strengths
by commutators of covariant derivatives that use
the fields $\csf_\eta$ and $\csf_\dt$.
Third, while in open superstrings $Q$ and $\eta$
are both derivations of the star algebra
and are on the same footing,
in heterotic string theory
$\eta$ is a derivation of all closed string products
but $Q$ is not.
Consequently, $Q$ and
$\eta$ are not
on the same footing and the fields
$\csf_Q$ and $\csf_\eta$ take inequivalent forms.
It would be exciting to find a completely
natural language for a geometrical formulation of the heterotic
string action.

There are several possible  generalizations of our results.
Open superstring field theory generalizes to open $N=2$ string
field theory \cite{Ooguri:1990ww}
by replacing $Q$ and $\eta$ with the generators
$G^+$ and $\tilde G^+$ from the $N=4$ topological description
of the open $N=2$ string \cite{Berkovits:1994vy,Berkovits:1998bt}.
Since the open bosonic string can be ``embedded''
in an open $N=2$ string \cite{Berkovits:1993xq},
the WZW action for open superstring field theory
can also be used to describe open bosonic string field theory.
Similarly, our heterotic results should generalize
to heterotic $N=2$ string field theory \cite{Ooguri:1991ie}
by replacing $Q_L + Q_R$ and $\eta_L$ with
the generators $G^+_L + Q_R$ and $\tilde G^+_L$ from the $N=4$
topological description of the heterotic $N=2$ string.
Since the closed bosonic string can be ``embedded''
in a heterotic $N=2$ string, our heterotic action
may be used to describe closed
bosonic string field theory.

One possible application of our action
is the study of tachyon condensation.
Just as the open superstring field theory action
was useful for testing Sen's conjectures
for open superstring tachyon condensation \cite{opentach},
it should be possible to use our heterotic string field theory action
to test conjectures
for closed superstring tachyon condensation~\cite{Adams:2001sv}.

Our results
indicate a close and intriguing relationship between structures
in open superstring field
theory and heterotic string field theory.  It is tempting to
speculate that such relationships exist because of
Type I/heterotic dualities. It
would also be interesting to generalize our construction to describe
the Ramond sector of heterotic string field theory.
Finally, it is still unknown how to construct
actions or equations of motion for Type II superstring field theory.
Since this theory has an additional fermionic direction $\eta_R$,
the construction might require new ingredients.

\medskip
\noindent{\bf Acknowledgments}\\
\noindent
The work of Y.O.~and B.Z.~was supported in part by
the DOE grant DF-FC02-94ER40818.
N.B. would like to thank Warren Siegel for useful discussions,
and CNPq grant 300256/94-9, Pronex grant
66.2002/1998-9, and FAPESP grant 99/12763-0 for partial financial
support.

\end{document}